\documentclass[]{aastex631}

\usepackage{savesym}
\savesymbol{tablenum}
\usepackage{siunitx}
\restoresymbol{SIX}{tablenum}

\usepackage{ragged2e}
\usepackage{rotating}
\usepackage{amsmath}
\usepackage{graphicx}
\usepackage{tablefootnote} 

\def\asec{\ifmmode ^{\prime\prime}\else$^{\prime\prime}$\fi}

\def\degs{\ifmmode ^{\circ}\else$^{\circ}$\fi}
\def\amin{\ifmmode ^{\prime}\else$^{\prime}$\fi}
\def\asec{\ifmmode ^{\prime\prime}\else$^{\prime\prime}$\fi}
\def\farcs{\hbox{$.\!\!^{\prime\prime}$}}  
\def\h{\hbox{$^{\rm h}$}}
\def\m{\hbox{$^{\rm m}$}}
\def\s{\hbox{$^{\rm s}$}}

\def\degs{\ifmmode ^{\circ}\else$^{\circ}$\fi}
\def\amin{\ifmmode ^{\prime}\else$^{\prime}$\fi}

\def\EE#1{\times 10^{#1}}

\def\cm{\mbox{\,cm}}

\def\cm3{\mbox{\,cm$^{-3}$}}
\def\kms{\mbox{\,km~s$^{-1}$}}

\def\ergs{\mbox{\,erg~s$^{-1}$}}

\def\kms{\mbox{\,km s$^{-1}$}}

\def\s{\mbox{\,s}}

\def\muJ{\mbox{$\mu$Jy}}
\def\lsim{\!\!\!\phantom{\le}\smash{\buildrel{}\over
 {\lower2.5dd\hbox{$\buildrel{\lower2dd\hbox{$\displaystyle<$}}\over
                                 \sim$}}}\,\,}
\def\gsim{\!\!\!\phantom{\ge}\smash{\buildrel{}\over
{\lower2.5dd\hbox{$\buildrel{\lower2dd\hbox{$\displaystyle>$}}\over
                               \sim$}}}\,\,}

\begin{document}

\title{SN 1885A and supernova remnants in the centre of M31 with LOFAR \footnote{Submitted on July 9, 2024; Accepted on October 18, 2024}}

\author[0000-0001-9896-6994]{Deepika Venkattu}
\affiliation{The Oskar Klein Centre, Department of Astronomy, Stockholm University, AlbaNova, SE-10691 Stockholm, Sweden \\}
\correspondingauthor{Deepika Venkattu}
\email{deepika.venkattu@astro.su.se}

\author[0000-0002-3664-8082]{Peter Lundqvist}
\affiliation{The Oskar Klein Centre, Department of Astronomy, Stockholm University, AlbaNova, SE-10691 Stockholm, Sweden \\}

\author[0000-0001-5654-0266]{Miguel Pérez Torres}
\affiliation{Instituto de Astrofísica de Andalucía (IAA-CSIC), Glorieta de la Astronomía, s/n, E-18008 Granada, Spain \\
}
\affiliation{School of Sciences, European University Cyprus, Diogenes street, Engomi, 1516 Nicosia, Cyprus\\}

\author[0000-0003-2312-3508]{Etienne Bonnassieux}
\affiliation{Julius-Maximilians-Universität Würzburg, Fakultät für Physik und Astronomie, Institut für Theoretische Physik und Astrophysik, Lehrstuhl für Astronomie, Emil-Fischer-Str. 31, D-97074 Würzburg, Germany\\}
\affiliation{LESIA, Observatoire de Paris, PSL University, CNRS, Sorbonne University,
Univ. Paris Diderot, Sorbonne Paris Cité, 5 place Jules Janssen, 92195 Meudon, France\\}
\affiliation{INAF, Istituto di Radioastronomia, Via Piero Gobetti, 101, 40129 Bologna BO, Italy\\}

\author{Cyril Tasse }
\affiliation{GEPI \& ORN, Observatoire de Paris, Université PSL, CNRS, 5 Place Jules Janssen, 92190 Meudon, France \\}
\affiliation{Department of Physics \& Electronics, Rhodes University, PO Box 94, Grahamstown, 6140, South Africa\\}

\author{Anne-Laure Melchior}
\affiliation{LERMA, Observatoire de Paris, Sorbonne University, PSL University, CNRS,
75014, Paris, France\\}

\author[0000-0003-2658-7893]{Francoise Combes}
\affiliation{Observatoire de Paris, LERMA, Collège de France, CNRS, PSL University, Sorbonne University, 75014, Paris, France}

\begin{abstract}
We present the first LOFAR image of the centre of M31 at a frequency of 150 MHz. We clearly detect three supernova remnants, which, along with archival VLA data at 3 GHz and other published radio and X-ray data allows us to characterize them in detail. Our observations also allow us to obtain upper limits the historical SN 1885A which is undetected even at a low frequency of 150 MHz. From analytical modelling we find that SN 1885A will stay in its free-expansion phase for at least another couple of centuries. We find an upper limit of $n_{\rm H}~\lsim 0.04$~cm$^{-3}$ for the interstellar medium of SN 1885A, and that the SN ejecta density is not shallower than $\propto r^{-9}$ (on average). From the $2.6\sigma$ tentative detection in X-ray, our analysis shows that non-thermal emission is expected to dominate the SN 1885A emission. Comparing our results with those on G1.9+0.3, we find that it is likely that the asymmetries in G1.9+0.3 make it a more efficient radio and X-ray emitter than SN 1885A. For Braun 80, 95 and 101, the other remnants in this region, we estimate ages of 5200, 8100, and 13\,100 years, and shock speeds of 1150, 880, and 660 \kms, respectively. Based on this, the supernova rate in the central 0.5~kpc $\times$ 0.6~kpc of M31 is at least one per $\sim 3000~{\rm yr}$. We estimate radio spectral indices of \textminus 0.66$\pm$0.05, \textminus 0.37$\pm$0.03 and \textminus 0.50$\pm$0.03 for the remnants, respectively, which match fairly well with previous studies.
\end{abstract}


\section{Introduction} \label{sec:intro}
M31, as the nearest spiral galaxy, 785 kpc away \citep{2005MNRAS.356..979M}, has been well-studied across multiple wavelengths. In the radio, dedicated surveys of M31 have been undertaken such as 36W \citep{1984A&AS...56..245B}, 37W \citep{1985A&AS...61..451W}, \cite{1990ApJS...72..761B} and \cite{2004ApJS..155...89G} surveys, apart from VLBI studies of parts of it (eg., \citealt{2013ApJ...768...12M}). These surveys have helped to build a census of all the radio sources in the galaxy, to understand the galaxy dynamics and also study individual sources that inform us about the past massive star population (supernova remnants, pulsars, pulsar wind nebulae) and current massive star population (H {\sc ii} regions) in M31.

Studying supernovae (SNe) and supernova remnants (SNRs) at radio wavelengths presents a picture of the synchrotron emission from these sources, helping us understand the interaction of the expanding ejecta with the surrounding circumstellar/interstellar matter and the nature of the progenitor stars. To study spatially resolved SNRs in great detail, galactic remnants with their large angular sizes on the sky are a natural target. However limitations include confusion and unreliable distance estimates. Studying SNRs in nearby galaxies is a good solution and M31 in the Local Group is one of these preferred targets, as it is 785 kpc away and is a spiral galaxy like the Milky Way. Many SNR studies in M31 have been performed in the optical (e.g., \citealt{1980A&AS...40...67D}; \citealt{1993A&AS...98..327B}; \citealt{1995A&AS..114..215M}) and a few in X-ray and radio as well (e.g., \citealt{1984MNRAS.206..351D}; \citealt{2001A&A...373...63S}; \citealt{2014SerAJ.189...15G}).

SN 1885A, discovered on 20th August 1885, was the first extragalactic supernova observed. Although it was also one of the brightest observed extragalactic supernovae, as a subluminous Type Ia supernova, it was never detected in the radio. \cite{1985ApJ...295..287D} present a centennial review of SN 1885A, providing a definitive light curve and reexamining colour and spectral observations pointing to its thermonuclear origins. There have been attempts in the past few decades to study the expanding ejecta of SN 1885A, and even though it has been more than a century after explosion, emission from the remnant at the site of SN 1885A has not been detected at any wavelength (e.g., \citealt{1989ApJ...341L..55F}; \citealt{1992ApJ...390L...9C}; \citealt{2019ApJ...872..191S}). The expanding ejecta, however, has been imaged in resonance-line \textit{absorption} in the optical and UV wavelengths \citep{1989ApJ...341L..55F, 2015ApJ...804..140F, 2017ApJ...848..130F}. Along with SNR G1.9+0.3 in our Galaxy, SN 1885A is one of two likely Type Ia SNe
that fill the gap in observations for our understanding of the SN--SNR transition phase.

For the first time, we present a LOw Frequency ARray (LOFAR; \citealt{2013A&A...556A...2V}) image of the centre of M31 and present fluxes for the three other SNRs in the region, and an upper limit for SN 1885A. We do not discuss M31* in this work, as, along with a sub-arcsecond resolution study of M31, it is discussed in detail in Bonnassieux et al (in prep). We focus only on the centre of M31 covering three SNRs, and SN 1885A in this work. Along with the LOFAR image, we include a Karl G. Jansky Very Large Array (VLA) image of the same region, and include fluxes from other radio and X-ray studies of these four sources. Following the literature, in this work, we refer to the three other SNRs as Braun 80, 95 and 101, as enumerated in the 1.4 GHz observations of \cite{1990ApJS...72..761B}.

In Section \ref{sec:obs} we present the observations and source detection with LOFAR and VLA. In Section \ref{sec:1885A} we present radio modelling of SN 1885A in the context of the upper limits provided by our observations and the 6.2 GHz VLA image \citep{2019ApJ...872..191S}. We discuss the X-ray upper limits in Section \ref{sec:x-rays} and compare our results for SN 1885A with G1.9+0.3 in Section \ref{g1.9}. Section \ref{sec:remnants} presents a discussion on morphology and spectra of the remnants Braun 80, 95 and 101 from radio and X-ray data available till date.

\section{Observations} \label{sec:obs}
\subsection{LOFAR}
LOFAR data are obtained from project LC10\_014 (PI: Dr. Anne-Laure Melchior) observed on 17th September 2018 (pointing P004+41). The observation using the High Band Antenna (HBA; 120-240 MHz) has a 10 minute observation of a bright flux density calibrator (3C 196) after the 8 hour on-source observation. The data were recorded with 1 second sampling and in channels of 12.21 kHz width (16 channels per subband, 243 subbands). Data were reduced and imaged using the standard LINC\footnote{\url{https://git.astron.nl/RD/LINC}} pipelines for direction-independent processing of the Dutch stations (formerly PREFACTOR; \citealt{2019A&A...622A...5D}) followed by direction-dependent calibration and imaging using DDFacet (\citealt{2023ascl.soft05008T}; \citealt{2018A&A...611A..87T}; \citealt{2015MNRAS.449.2668S}). The image, centred at 145 MHz, has a beam size of \SI{6}{\arcsecond} and rms of 0.1 mJy/beam.

\subsection{VLA}
VLA data are obtained from project 22A-071 (PI: Dr. Zhiyuan Li) observed on 27th June 2022. The C-band image centred at 3 GHz has a beam size of \SI{0.5}{\arcsecond} and rms of 3 \muJ/beam and was taken when the array was in A configuration. Data were processed through the VLA Calibration Pipeline with the flux calibrator 3C48 and the complex gain calibrator J0038+4137 and imaged with a briggs weighting of 0.5. 

In addition to the data at 150 MHz and 3 GHz, we use previously published data in radio and X-ray from \cite{1990ApJS...72..761B}, \cite{2001AIPC..565..433S}, \cite{2003ApJ...590L..21K}, \cite{2019ApJ...872..191S} and \cite{2013A&A...555A..65H}, as given in Tables \ref{tab:flux}, \ref{tab:Xray_flux_snrs} and \ref{tab:Xray_flux_85a}.

\subsection{Source detection} \label{source}

In the LOFAR image, the remnants Braun 80, 95 and 101 remain barely resolved, show no detailed structure but are slightly bigger than the image beam. For this image, we use the default parameters for point sources in \textsc{PyBDSF}. There is significant extended emission from this central region in the LOFAR image, and no clear emission that can be isolated as coming from SN 1885A itself. For a \SI{6}{\arcsecond} region centred on SN 1885A, we get a flux density of 0.797 mJy which we treat as an upper limit for the source at 150 MHz. 

For the VLA S-band image, where the remnants are resolved, we use the \textsc{IMFIT} function in \textsc{CASA}. The fit parameters for both images are given in Table \ref{tab:flux}.
Although we do not discuss them further, we also add other point sources seen in the two images in Table \ref{tab:source_det} characterised with \textsc{PyBDSF}. 

\begin{deluxetable}{lcccccccc}
\tabletypesize{\scriptsize}
\tablewidth{0pt}  
\centering
\caption{SNRs in the M31 core - radio fluxes }
\label{tab:flux}
\tablehead{
\colhead{SNR} & \colhead{RA}& \colhead{Dec} & \colhead{Central Frequency} & \colhead{Integrated Flux Density} & \colhead{Major} & \colhead{Minor} & \colhead{PA} & \colhead{Source}\\
\colhead{} & \colhead{}& \colhead{} & \colhead{($\rm GHz$)} & \colhead{($\rm mJy$)} & \colhead{($\rm arcsec$)} & \colhead{($\rm arcsec$)} & \colhead{($\rm deg$)} & \colhead{}\\
} 
\startdata 
Braun 80  & 00h42m40.26s & 41d15m52.1s & 0.145 & $8.3\pm1.4$ & 14.11$\pm$2.14 & 11.02$\pm$1.52 & 97.85$\pm$28.68 & This work\\
          &              &             & 0.145 & 6.5$\pm$1.5 & 10.64$\pm$1.94  & 8.79$\pm$1.4 & 121.7$\pm$41.4 & LoTSS$^{\rm a}$\\
          &              &             & 1.4 & 3.12$\pm$0.46 &  21.17$\pm$1.54 & 15.95$\pm$1.11 & 156.0$\pm$15.2 & [1] \\    
          &              &             & 3 & 0.62$\pm$0.10 & 6.17$\pm$1.02 & 5.55$\pm$0.92 & 83$\pm$63 & This work\\ 
          &              &             & 8.4 & 0.53$\pm$0.03$^{\rm b}$ & 8.0 & 8.0 & - &  [2]\\
\hline          
Braun 95 & 00h42m47.82s & 41d15m25.07s & 0.145 & $11.27\pm1.19$   & 10.50$\pm$0.86 & 10.09$\pm$0.81 & 162.05$\pm$85.32 & This work\\
          &              &             & 0.145 & 10.7$\pm$1.7 & 11.68$\pm$1.49 & 9.29$\pm$1.05 & 107.19$\pm$24.6 & LoTSS$^{\rm a}$\\
         &              &             & 1.4 & 2.48$\pm$0.19 &  9.79$\pm$0.25 & 7.17$\pm$0.27 & 108.3$\pm$5.1 & [1]\\          
          &              &             & 3 & 1.03$\pm$0.11 & 6.76$\pm$0.73 & 5.92$\pm$0.64 & 142$\pm$33 & This work\\ 
         &              &             & 8.4 & 1.41$\pm$0.05$^{\rm b}$ & 9.6 & 8.0 & 137 &  [2]\\ 
\hline          
Braun 101 & 00h42m50.41s & 41d15m56.4s & 0.145 & $9.74\pm0.9$ & 10.34$\pm$0.79 & 8.55$\pm$0.58 & 146.05$\pm$17.54 & This work\\
          &              &             & 0.145 & 14.5$\pm$2.0 & 13.9$\pm$1.68 & 10.99$\pm$1.2 & 0.12$\pm$23.2 & LoTSS$^{\rm a}$\\
          &              &             & 1.4 & $2.97\pm0.23$ &  $10.93\pm0.34$ & $7.77\pm0.53$ & $133.9\pm6.9$ & [1]\\  
          &              &             & 3 & 1.14$\pm$0.10 & 6.24$\pm$0.55 & 5.52$\pm$0.49 & 169$\pm$29 & This work\\
          &              &             & 4.9 & 1.13$\pm$0.05$^{\rm c}$ & 11.6 & 8.4 & - &  [3]\\          
          &              &             & 8.4 & 1.29$\pm$0.05$^{\rm b}$ & 11.6 & 8.4 & 143 &  [2]\\      
\hline
\enddata

\tablecomments{\begin{flushleft}
    \hspace{5mm}  $^{\rm a}$ \url{https://vo.astron.nl/lotss_dr2/q/src_cone/form}\\
    \hspace{5mm}  $^{\rm b}$ We add 3\% error in addition to the 3$\sigma$ from rms noise of 4 \muJ \ reported in [2]\\
    \hspace{5mm}  $^{\rm c}$ We add 3\% error in addition to the 3$\sigma$ from rms noise of 5 \muJ \ reported in [3] 
\end{flushleft}
References: [1] \cite{1990ApJS...72..761B}; [2] \cite{2001AIPC..565..433S}; [3] \cite{2003ApJ...590L..21K}.}
\end{deluxetable}

\begin{table}
\centering
\caption{SNRs in the M31 core - X-ray data}
\label{tab:Xray_flux_snrs}
\begin{tabular}{cccc} 
\hline
 SNR    &   X-ray energy band              & Luminosity             &  Source     \\
        &    keV                           & $\times 10^{35}$~erg s$^{-1}$ &       \\      
\hline
Braun 80 &  0.2 - 10 & 3.2 & Estimated using [1]    \\
\hline
Braun 95 &   0.2 - 10 & 5.3 & Estimated using [1] \\
         &   0.3 - 7  & $8^{+7.0}_{-2.2}$  & [2] \\
\hline
Braun 101 & 0.2 - 10  & $6.2\pm0.49$ & [1] \\
         &   0.3 - 7  & $4.4^{+1.3}_{-2.2}$  & [2] \\
         &  0.2 - 4.5  & $6.0\pm0.51$ & Source 1050 in [3] \\
         &  0.35 - 2.0 & $12$      & XMMM31 J004250.47+411556.7 in [4] \\
\hline
\end{tabular}
\tablecomments{References: [1] \cite{2013A&A...555A..65H}; [2] \cite{2003ApJ...590L..21K}; [3] \cite{2011A&A...534A..55S}; [4] \cite{2012A&A...544A.144S} } 
\end{table}

\begin{table}
\centering
\caption{SN 1885A - Radio and X-ray limits}
\label{tab:Xray_flux_85a}
\begin{tabular}{cccccc} 
\hline
 SNR    & Age & Central Frequency/X-ray energy band   & Integrated Flux Density           & Luminosity             &  Source     \\
        & years & GHz/keV         &$\rm mJy$     & ~erg s$^{-1}$ Hz$^{-1}$ &       \\      
\hline
SN 1885A     & 133 & 0.15 GHz & $<$ 0.8 &  $<$ 3.46$\times 10^{23}$  & This work  \\
             & 127 & 6.2 GHz  & $<$ 0.01 & $<$ 8.43$\times 10^{21}$ & [1] \\
             & 123 & 0.2 - 10 keV & - & $<$ 2.0$\times 10^{34}$ & [2]\\
             & 123 & 0.3 - 7 keV & -  & $<$ 1.6$\times 10^{34}$ & Estimated using [2] and [3]  \\
\hline
\end{tabular}
\tablecomments{References: [1] \cite{2019ApJ...872..191S}; [2] \cite{2013A&A...555A..65H}; [3] \cite{2003ApJ...590L..21K} } 
\end{table}

\begin{deluxetable}{lccccccc}
\tabletypesize{\scriptsize}
\tablewidth{0pt}  
\centering
\caption{Other sources in the M31 core - radio fluxes from this work }
\label{tab:source_det}
\tablehead{
\colhead{Source} & \colhead{RA}& \colhead{Dec} & \colhead{Central Frequency} & \colhead{Integrated Flux Density} & \colhead{Major} & \colhead{Minor} & \colhead{PA} \\
\colhead{} & \colhead{}& \colhead{} & \colhead{($\rm GHz$)} & \colhead{($\rm mJy$)} & \colhead{($\rm arcsec$)} & \colhead{($\rm arcsec$)} & \colhead{($\rm deg$)} \\
} 
\startdata 
GLG 001/37W 142  & 00h42m45.21s &  41d15m05.58s  & 0.145 & 9.1$\pm$0.6 & 7.1$\pm$0.3 & 6.8$\pm$0.3 & 82.1$\pm$49.7 \\
                 &              &             & 3 & 1.2$\pm$0.01 & 0.55$\pm$0.001 & 0.53$\pm$0.001 & 68.6$\pm$2.3 \\ 
\hline          
M31N 2010-04a & 00h42m44.97s  &  41d15m09.47s & 0.145 & $20.7\pm2.9$   & 12.4$\pm$3.3 & 7.6$\pm$1.5 & 125.8$\pm$25.8 \\
              &              &             & 3 & 0.1$\pm$0.01 & 0.6$\pm$0.02 & 0.52$\pm$0.02 & 179.5$\pm$10.2 \\ 
\hline          
CXOU J004249.11+411456.6 & 00h42m49.12s & 41d14m56.48s & 3 & 0.04$\pm$0.01 & 0.76$\pm$0.13 & 0.74$\pm$0.13 & 145.7$\pm$0.01 \\
\hline
M31*                     & 00h42m44.32s & 41d16m08.43s & 3 & 0.028$\pm$0.006 & 0.71$\pm$0.11 & 0.52$\pm$0.06 & 178.2$\pm$21.3 \\
\hline
2CXO J004245.1+411659    & 00h42m45.17s & 41d16m59.16s & 3 & 0.04$\pm$0.008 & 0.8$\pm$0.1 & 0.6$\pm$0.07 & 109.6$\pm$23 \\ 
\hline
\enddata
\tablecomments{\begin{flushleft}
    \hspace{5mm}  For cross-identifications and references, the reader is referred to the NASA/IPAC Extragalactic Database\\
\end{flushleft}}
\end{deluxetable}

\begin{figure}[ht!]
\includegraphics[width=0.6\textwidth]{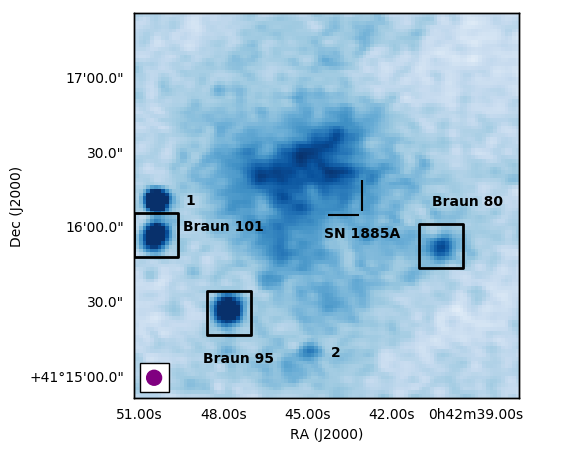}
\caption{LOFAR image of the centre of M31 at 150 MHz with Braun 80, 95 and 101 labelled and the position of SN 1885A marked. The bright source north of Braun 101 is GLG 001, and the source south-west of Braun 95 is M31N 2010-04a (marked 1 and 2 respectively; see also Table \ref{tab:source_det}). The restoring beam is shown in the bottom left corner. \label{fig:m31lofar}}
\end{figure}

\begin{figure}[ht!]
\includegraphics[width=0.6\textwidth]{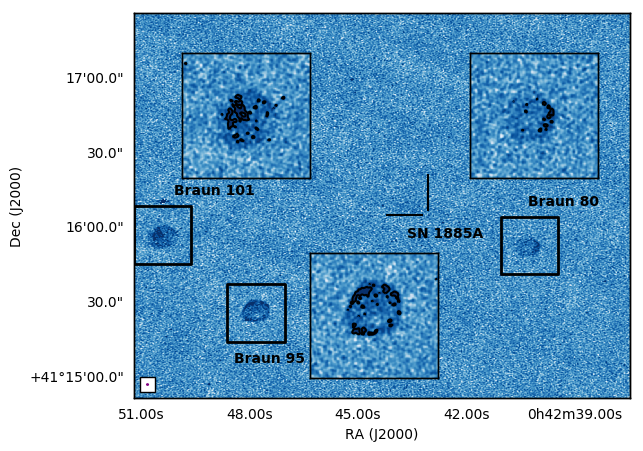}
\caption{VLA image of centre of M31 at 3 GHz with Braun 80, 95 and 101 labelled and the position of SN 1885A marked. The insets show the structure of the three remnants in greater detail. Bright point sources visible in this image are noted in Table \ref{tab:source_det}. The restoring beam is shown in the bottom left corner.  \label{fig:m31sband}}
\end{figure}

\section{Discussion}
\subsection{SN 1885A}
\label{sec:1885A}

Radio observations of SN 1885A have so far been unsuccessful in detecting any emission (e.g., \citealt{1967Natur.214.1190P}; \citealt{1984MNRAS.206..351D}; \citealt{1992ApJ...390L...9C}). \cite{2019ApJ...872..191S} obtain the deepest radio image at the site of SN 1885A by co-adding archival VLA observations between 4 and 8 GHz. They provide a 3$\sigma$ upper limit of $<$11.4 \muJ/beam at a central frequency of 6.2 GHz at 127 years. We note here that although we use the term `SN 1885A' for uniformity with literature, the source is in its remnant phase and the discussion that follows also concerns the remnant `SNR 1885A'.

Observations in the UV and optical by \cite{1989ApJ...341L..55F, 2015ApJ...804..140F, 2017ApJ...848..130F} show evidence for expanding debris at the position of SN 1885A from resonance-line absorption against the background emission from M31. 
The fastest material has a velocity of $\approx 13,400 \kms$ as revealed by \ion{Ca}{1} $\lambda 4227$ data from 2013 \citep{2017ApJ...848..130F}. 
\cite{2017ApJ...848..130F} compare the observationally inferred velocity structure of the debris with that of a subluminous Type Ia SN model named 5p02822.16 by \cite{2002ApJ...568..791H}, and find qualitative similarities, enough to suggest that SN 1885A was the result of an off-center delayed-detonation explosion. 
In particular, Ca is present in the model out to $\approx13\,000 \kms$. In the model there is also Si, S, and especially Mg present out to $\approx18\,000 \kms$, with little or no mixing between Ca and Mg. 
\cite{2017ApJ...848..130F} did not find any observational evidence of Mg at velocities $\gsim 13\,000 \kms$ which could signal a difference between the model and the observations. Or, it could be that ejecta with velocity originally greater than $\approx13\,400 \kms$ could have been swept up by the reverse shock created by SN ejecta--ISM interaction, by the time of the observations in 2013 (at the age of 128 yrs for SN 1885A). We suppose the latter to be the case.

To model the SN ejecta, we assume that the ejecta are spherically symmetric and expand in a homologous fashion, i.e., $V(r,t) = r/t$, where $V$ is the velocity, $r$ the radius, and $t$ the time since explosion. We will use cgs units unless otherwise stated. We further presume that the ejecta density structure can be approximated by two power-laws, where the inner structure is characterized by the density profile $\rho_i (V,t)\propto V^{-a}t^{-3}$ and the outer structure by $\rho_o (V,t) \propto V^{-n}t^{-3}$. The break in density slope between these two parts of the ejecta occurs at the velocity $V_{\rm b}$, which can be calculated from integrating the density and kinetic energy profiles across the ejecta, to become
\begin{equation}
	V_{\rm b}= 10\,030~\sqrt {\frac{E_{51}}{M}  \frac{\left(n-5\right) \left(5-a\right)} {\left(n-3\right)\left(3-a\right)}}~{\rm km~s}^{-1}.
\label{eq:Vb}
\end{equation}
Here $E_{51}$ is the total kinetic energy of the ejecta in $10^{51}$ erg, and $M$ the total ejecta mass in solar masses (see also \citealt{1994ApJ...420..268C}). Typical parameters for a Type Ia SN explosion are $E_{51} = 1$  and $M = 1.4$ M$_{\odot}$, and an oft-used value for $n$ is $n=10$ (e.g., \citealt{1999ApJS..120..299T}), although $n=12-14$ have also been used for the outermost ejecta (e.g., \citealt{2017ApJ...842...17K}). For the parameter $a$, the model 5p02822.16 indicates a value of $0-1$ (see Fig. 15 of \citealt{2017ApJ...848..130F}), and the combination $E_{51} = 1$, $M = 1.4$ M$_{\odot}$, $n=10$, and $a=0~(1)$ gives $V_{\rm b} \approx 9\,250~(10\,100) \kms$. In the model 5p02822.16 there is indeed a break in density at $\sim 9\,000 \kms$, but the density slope exterior to this velocity is shallower than $n=10$ out to $\sim 23\,000 \kms$. It is more like the $n=7$ scenario discussed by  \cite{1982ApJ...258..790C} for Type I SNe. However, a steeper density slope could be expected for the ejecta with the highest velocities (see, e.g., \citealt{1998ApJ...497..807D} and \citealt{2017ApJ...842...17K}). We use the $a$-values 0 and 1, and $n$-values between 7 and 12 as approximations to the full density profile. \cite{2019ApJ...872..191S} used $a=0$ and $n=10$ in their models for SN 1885A. Our models M1--M8 are summarized in Table \ref{tab:an_models}.

The fact that SN 1885A was subluminous (e.g., \citealt{1885Obs.....8..321M,1886MmSSI..14..144M}; \citealt{1885Natur..32..465H}; \citealt{1988ApJ...331L.109C}) may argue for $E_{51} \lsim 1$, but as there is no clear correlation between explosion energy and production of $^{56}{\rm Ni}$ in the delayed-detonation scenario (cf. \citealt{2002ApJ...568..791H}), we select a value of $E$ for each combination of $a$ and $n$ so that the break in density structure occurs at $\approx 9\,000 \kms$. In the 5p02822.16 model by \cite{2002ApJ...568..791H}, $E_{51}\approx1.106$ (P. H\"oflich, private communication), which is close to that in our models M2 and M8. The ejecta mass is fixed at $M = 1.4$ M$_{\odot}$.

\begin{table*}
\centering
\caption{Model parameters for SN 1885A} 
\label{tab:an_models}
\begin{tabular}{lcrclclcrrrcc} 
\hline
Model        & $n$ & $a$ & $E$                   & $n_{\rm H}$  & $t_{\rm b}$  & $M_{1,128}$  & $R_{1,128}$  & $V_{1,128}$ & $V_{2,128}$ & $B_{128}^{\rm a}$  & $f_{\rm rel,128}^{\rm b}$ & $L(150~{\rm MHz})_{133}^{\rm a,b}$ \\
                  &    &    & $10^{51}$ erg & cm$^{-3}$     & yrs               &  M$_{\odot}$  & pc                  & $\kms$\         & $\kms$         & $\mu$G      & $\times 10^{-4}$  & $\times 10^{22}$~erg s$^{-1}$ Hz$^{-1}$ \\      
\hline
M1             & 7  & 0  &         1.35              &      0.155        &  324   &  0.244   &  2.22  &  9\,670 & 4\,310         & 292 & 1.28 & 232 \\
M2             & 7  & 1  &         1.13        &            0.121           & 324            & 0.190 & 2.22  &  9\,670  & 4\,310 & 258 & 1.28 & 147\\
M3             & 10  & 0  &       0.95               &     0.0188       & 482      &  0.024  &   2.05  &  10\,980  & 3\,020 & 115 & 1.66 & 4.81 \\
M4             & 10  & 1  &        0.79         &          0.0140   &      482  &  0.017  & 2.05   &  10\,980  & 3\,020 & 99 & 1.66 & 2.80 \\
M5             & 12  & 0  &       0.87            &        0.0051      &  629       &  0.0061  & 2.02    &  11\,570 & 2\,510 & 63 & 1.87 & 0.48 \\
M6             & 12  & 1  &        0.72        &           0.0037   &     629    &  0.0044  & 2.02  & 11\,570 & 2\,510 & 54 & 1.87 & 0.27 \\
M7             & 8  & 0  &       1.13            &        0.0722      &  370       &  0.101  & 2.13    &  10\,170 & 3\,760 & 210 & 1.42 & 54.7 \\
M8             & 9  & 0  &        1.02        &           0.0352   &     422    &  0.046  & 2.08  & 10\,610 & 3\,350 & 153 & 1.54 & 14.8 \\
\hline
\end{tabular}
\RaggedRight
\tablecomments{\begin{flushleft}
    \hspace{5mm}  $^{\rm a}$Evaluated for $\epsilon_B = 0.01$.\\
    \hspace{5mm}  $^{\rm b}$Evaluated for $\epsilon_{\rm rel} = 0.001$ and $p=2.3$.\\
\end{flushleft}
}
\end{table*}

As the SN ejecta expand, they will interact with the ISM. For a helium-to-hydrogen ratio (by number) of $n_{\rm He}/n_{\rm H} = 0.1$, the density of the ISM is $\rho_{\rm ISM} = 1.4 m_{\rm p} n_{\rm H}$, where $n_{\rm H}$ is the hydrogen density in cm$^{-3}$.
The interaction between the ejecta and the ISM creates two shock waves, a forward one moving into the ISM, and a reverse shock moving into the ejecta. Since we assume a power-law density structure of the ejecta, and that the ISM density has a constant value, the shock structure can be obtained from the similarity solutions of \cite{1982ApJ...258..790C}, as long as the velocity of the outermost ejecta $V_{\rm ej,max} \geq V_{\rm b}$ and $n > 5$. Time $t_{\rm b}$ occurs when $V_{\rm ej,max} = V_{\rm b}$ and is simply $t_{\rm b} = R_2(t_{\rm b}) /V_{\rm b}$, where $R_2(t_{\rm b})$ is the reverse shock radius at time $t_{\rm b}$. As long as $t \leq t_{\rm b}$, $R_2$ and $R_1$ (the radius of the forward shock) are related through $R_2 = R_1 (R_2/R_{\rm c}) (R_1/R_{\rm c})^{-1}$ where the ratios of $R_1/R_{\rm c}$ and $R_2/R_{\rm c}$ are tabulated in \cite{1982ApJ...258..790C} for various values of $n$. Here, $R_{\rm c}$ is the radius of the contact discontinuity between shocked ejecta and shocked ISM. $R_1$ can be written as
\begin{equation}
	R_1(t)= R_{1,128}~\left( \frac{t_{\rm yr}} {128} \right)^{(n-3)/n},
\label{eq:R1}
\end{equation}
where $R_{1,128}$ is the outer shock radius at 128 yrs, i.e., at the time of the observations by \cite{2017ApJ...848..130F}, and $t_{\rm yr}$ is the time $t$ since explosion in years. 

We make use of the \ion{Ca}{1} line analysis of \cite{2017ApJ...848..130F} which shows that \ion{Ca}{1} is present all the way out to $13\,400 \kms$. We therefore select $V_{\rm ej,max}({\rm 128~yrs}) =13\,400 \kms$ as the maximum ejecta velocity at 128 years, corresponding to the reverse shock radius $R_{2,128} \approx 1.75$~pc (or a projected radius of $0\farcs46$ at a distance of 785 kpc). $R_{1,128}$ is then given by $R_{1,128} = R_{2,128} (R_1/R_{\rm c}) (R_2/R_{\rm c})^{-1}$, and is listed in Table \ref{tab:an_models} for our eight models. The velocity of the forward shock at 128 years is given by $V_{1,128} = V_{\rm ej,max}({\rm 128~yrs}) (R_1/R_{\rm c}) (R_2/R_{\rm c})^{-1}(n-3)n^{-1}$, and is also listed in Table \ref{tab:an_models}.  In their models, \cite{2019ApJ...872..191S} assumed  $12\,500 \kms$ as a lower limit to the velocity of the outer shock, which (for $n=10$) would require $V_{\rm ej,max}({\rm 128~yrs}) \approx 15\,250 \kms$, i.e., faster ejecta than observed by \cite{2017ApJ...848..130F}. For the sake of completeness we also list in Table \ref{tab:an_models} the velocity of the reverse shock, where we made use of the ratio of fluid velocities of shocked ejecta and shocked ISM at their respective shock fronts as given by \cite{1982ApJ...258..790C}.

The amount of swept up ejecta by the reverse shock at $t_b$ is $M_2(t_{\rm b}) = (3-a) (n-a)^{-1} M$, and at 128 yrs, the corresponding mass, $M_{2,128}$, is a fraction $(0.9/1.34)^{(n-3)}$ smaller. The mass of ISM swept up by the forward shock is given by $M_{1,128} = M_{2,128} (M_2/M_1)^{-1}$, where the $M_2/M_1$ ratio is taken from \cite{1982ApJ...258..790C}. The density of the unshocked ISM is then found from
\begin{equation}
	n_{\rm H} = 0.171 \frac{M_{1,128}} {m_{\rm p} R_{1,128}^3}~{\rm cm}^{-3},
\label{eq:rho}
\end{equation}
and $t_{\rm b} = 128~(1.34/0.9)^{n/3}$ yrs. The general dependence on  $t_{\rm b}$ is $t_{\rm b} \propto \rho_{\rm ISM}^{-1/3} M^{5/6} E^{-1/2}$.
The case with $a=0$, $n=7$, $E_{51}=1$, $M=1.4$ M$_{\odot}$, and $n_{\rm H} = 1$ cm$^{-3}$, for which $t_{\rm b} \approx 204$ yrs, was discussed in \cite{1982ApJ...258..790C}. For n=10, and other parameters being the same as in \cite{1982ApJ...258..790C}, $t_{\rm b} \approx 232$ yrs. This case was studied by \cite{2019ApJ...872..191S} for SN 1885A. We find larger values for $t_{\rm b}$ (Table \ref{tab:an_models}), and it seems the remnant will remain in its free expansion phase for at least another $\sim 2$ centuries.

\subsubsection{Radio} \label{radio_85}

To model the radio emission from the interaction between the ejecta and the ISM, we assume that the fraction $\epsilon_B$ of the forward shock energy density $\rho_{\rm ISM} V_1^2$, goes into magnetic field energy density, $u_B$, and the fraction $\epsilon_{\rm rel}$ goes into relativistic electron energy density, $u_{\rm rel}$. 

For the relativistic electrons, we assume a power law distribution of the electron energies, $n(\varepsilon) = N_0\varepsilon^{-p}$, 
where $\varepsilon=\gamma m_e c^2$ is the energy of the electrons and $\gamma$ is the Lorentz factor. From this one finds that
\begin{equation}
	N_0 = \epsilon_{\rm rel} \left(p-2\right) \rho_{\rm ISM} V_1^2 \left(\gamma_{\rm min} m_e c^2\right)^{p-2},
\label{eq:N0}
\end{equation}
where $\gamma_{\rm min} m_e c^2$ is the minimum energy of electrons contributing to the synchrotron emission. 
The parameter $N_0$ can also be expressed as 
\begin{equation}
	N_0 = f_{\rm rel}\left(p-1\right) n_{e,1} \left(\gamma_{\rm min} m_e c^2\right)^{p-1},
\label{eq:N0_2}
\end{equation}
where $n_{e,1}$ is the postshock electron density at the forward shock and $f_{\rm rel} = n_{\rm rel,1}/n_{e,1}$. Here
$n_{\rm rel,1}$ is the density of the postshock electrons with $\gamma \geq \gamma_{\rm min}$ 
with the condition that $\gamma_{\rm min} \geq 1$. $\eta$ is the shock compression factor (taken to be $\eta = 4$) and 
\begin{equation}
	\mu_e = \frac{1+ 4 \frac{n_{\rm He}} {n_{\rm H}}} {1+ 2 \frac{n_{\rm He}} {n_{\rm H}}}.  
\label{eq:mu_e}
\end{equation} 
Equations~\ref{eq:N0} and~\ref{eq:N0_2} give
\begin{equation}
	\gamma_{\rm min} = \left(\frac{p-2}{p-1}\right) \left(\frac{m_p}{m_e}\right) \left(\frac{\epsilon_{\rm rel} \mu_e}{\eta f_{\rm rel}}\right) \left(\frac{V_1}{c}\right)^2,
\label{eq:g_min}
\end{equation}
but since $f_{\rm rel}$ is unknown, we have chosen to initially put $f_{\rm rel} = 1$ in Eqn~\ref{eq:g_min}
and call this $\gamma$-value $\gamma_{\rm min_1}$. We then estimate $\gamma_{\rm min}$ from
\begin{equation}
    \gamma_{\rm min} = {\rm max} (1,\gamma_{\rm min_1}),
\label{eq:g_min_1}
\end{equation}
and estimate $f_{\rm rel}$ from
\begin{equation}
	f_{\rm rel} = \frac{\gamma_{\rm min_1}}{\gamma_{\rm min}}
\label{eq:f_rel}
\end{equation}
(see also \citealt{2023ApJ...952...24H}).

We have used $\epsilon_{\rm rel}=0.001$, which is close to the geometric mean of $9.5\EE{-4}$ for the young Type Ia SNRs Tycho, Kepler and G1.9+0.3 \citep{2021ApJ...917...55R}. For this low value of $\epsilon_{\rm rel}$, $\gamma_{\rm min} = 1$ (cf. Eqn~\ref{eq:g_min}) and $f_{\rm rel} < 1$ (see also \citealt{2023ApJ...952...24H}). 
Table \ref{tab:an_models} lists $f_{\rm rel}$ at 128 years for our eight models. Note that $f_{\rm rel}$ does not explicitly depend on density. For $p$ we have selected $p=2.3$ which is within the range $p \in [2.2,2.5]$ typical for SNRs (e.g., \citealt{2019JApA...40...36G}), but we note that for the specific case of G1.9+0.3, $p\approx2.6$ \citep[][see also Sect. 3.1.3]{2020MNRAS.492.2606L}.

For the postshock magnetic field energy density, $B = (8 \pi u_B)^{1/2}$, and in Table \ref{tab:an_models} we list $B$ at 128 years for our models assuming $\epsilon_B = 0.01$. The values of $B$ for models M1, M2, and M7 are similar to those of Kepler, Tycho, and G1.9+0.3 which have a geometric mean of $229~\mu G$ \citep{2021ApJ...917...55R}. The values for RX~J1713.7-3946, RCW~86 and SN~1006 are $\sim 100~\mu G$ and thus close to the values for our models M3 and M4. It therefore seems that $\epsilon_B = 0.01$ is of the right order, and it also agrees with the geometric mean of $\epsilon_B \approx 0.017$ for Tycho, Kepler and G1.9+0.3 \citep{2021ApJ...917...55R}. \cite{2019ApJ...872..191S} devised a complete analytical approach to estimate the evolution of $\epsilon_B$, but since we are mainly interested in only one epoch (at $\sim 130$ years), and since the method does not reproduce observed radio emission for individual SNRs \citep{2022Univ....8..653L}, we have used a fixed value for $\epsilon_B$.

The intensity of optically thin synchrotron emission is $\propto \nu^{-\alpha}$, where $\alpha = (p-1)/2$. For $p=2.3$, $\alpha = 0.65$.
If we include synchrotron self-absorption (SSA), the radio luminosity $L_{\nu}$, can be written as, 
\begin{equation}
\begin{split}
		L_{\nu} = 4 \pi^2 R_1^2  \frac{c_5}{c_6} B^{-1/2} \left( 1-{\rm exp} \left[- \left(\frac{\nu}{\nu_1}\right)^{-(p+4)/2} \right] \right) \left( \frac{\nu}{2\nu_1} \right)^{5/2} \\
	{\rm erg}~{\rm s}^{-1}~{\rm Hz}^{-1}.
\end{split}
\label{eq:L_nu}
\end{equation}
Optical depth unity occurs at the frequency $\nu_1$ given by
\begin{equation}
	\nu_1 = 2 c_1 (s c_6 N_0)^{2/(p+4)} B^{(p+2)/(p+4)}~{\rm Hz}. 
\label{eq:nu_1}
\end{equation}
The constants $c_1$, $c_5$, and $c_6$ are from \cite{1970ranp.book.....P} and have the values (in cgs units) $6.27\EE{18}$, $9.68\EE{-24}$, and $8.1\EE{-41}$, respectively. $s$ is the thickness of a cylindrical slab such that 
\begin{equation}
	\pi R_1^2 s  = \frac{4 \pi R_1^3}{3} \left[1 - \left(\frac{R_2}{R_1}\right)^3 \right]~{\rm cm}^3.
\label{eq:s_thick}
\end{equation}
 The volume of synchrotron emission is assumed to be the entire volume between the reverse and forward shocks. 
 
 \begin{figure}
\includegraphics[width=0.7\textwidth]{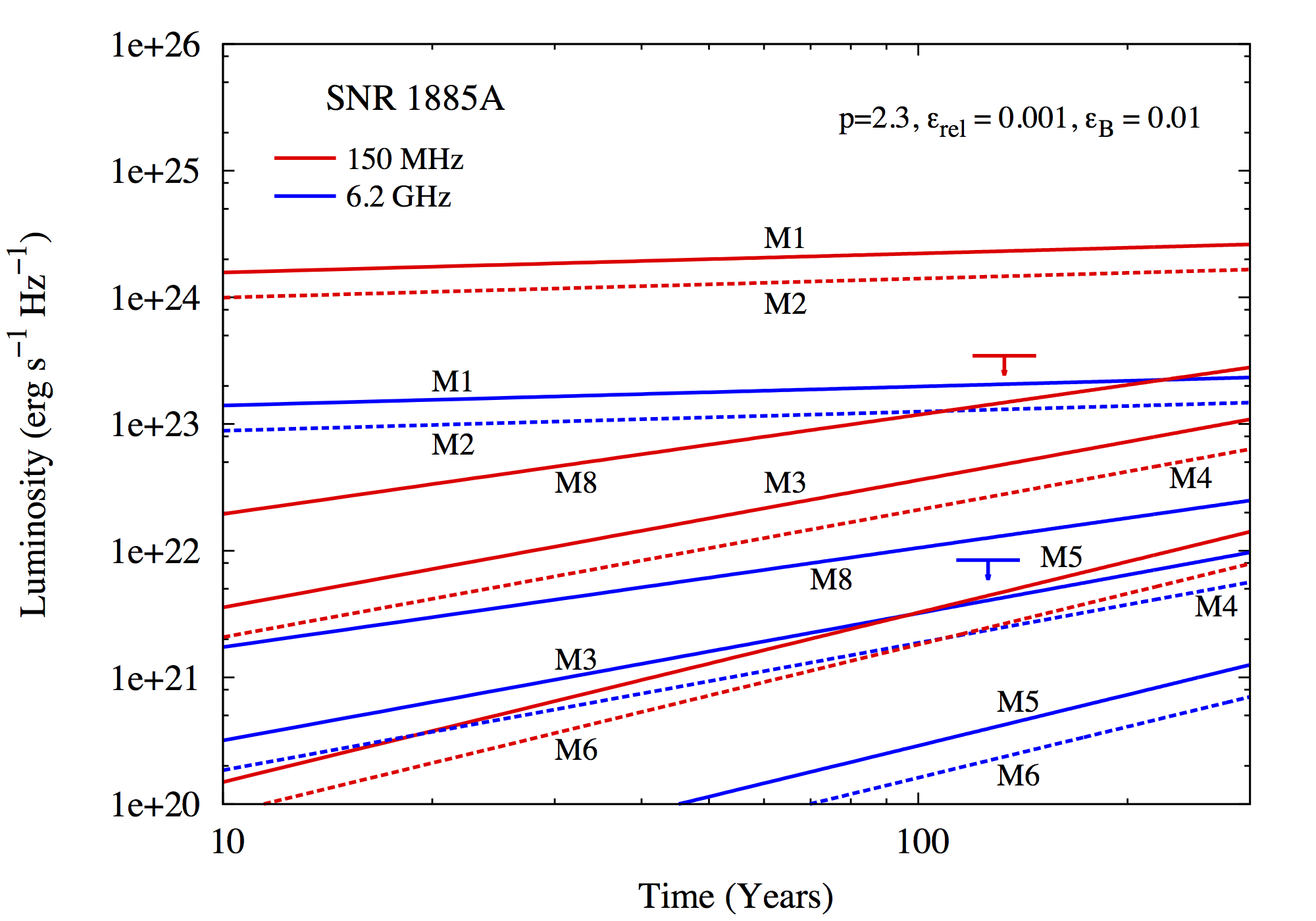}
\caption{SN 1885A at 150 MHz (red lines) and 6.2 GHz (blue). Observed upper limits at 127 years (for 6.2 GHz) and 133 years (for 150 MHz) are marked, as well as the modeled evolution for models M1--M6 and M8 in Table~\ref{tab:an_models}. For all models we have assumed the parameters $\epsilon_{rel}=0.001$, $\epsilon_B=0.01$, and $p=2.3$,}
\label{fig:1885Amod}
\end{figure}

The evolution of the modeled luminosities for 150 MHz and 6.2 GHz for models M1--M6 and M8 is shown in Fig.~\ref{fig:1885Amod}.
For all the models, $\nu_1 \ll  150 $~MHz even just one year after explosion, indicating optically thin emission, so SSA is unimportant for SN 1885A. The observed upper limits of $3.46\EE{23}$~erg s$^{-1}$ Hz$^{-1}$ at 150 MHz and 133 years, and $8.43\EE{21}$~erg s$^{-1}$ Hz$^{-1}$ at 6.2 GHz and 127 years, are marked in the figure.  The 150 MHz luminosities for the models are listed in Table \ref{tab:an_models}. The 6.2 GHz limit is the most constraining and SN 1885A would have been detected at 6.2 GHz if similar to models M1, M2, and M8, and at 150 MHz if similar to M1 and M2. For $p > 3$, the LOFAR data would have been more constraining than the 6.2 GHz data.

The modeled emission is sensitive to the parameter $n$, but insensitive to $a$. With the choices of $\epsilon_{\rm rel}=0.001$, $\epsilon_B=0.01$, $p=2.3$, 
and a structure similar to that of the explosion model 5p02822.16 \citep{2002ApJ...568..791H}, $n~\gsim 9$, $n_{\rm H}~\lsim 0.035$~cm$^{-3}$, and $E_{51}~\lsim 1$.
\cite{2019ApJ...872..191S} assumed $n=10$, a fixed value for $V_1$ at 128 years of $12\,500 \kms$, $p=2.2$, and $\epsilon_{\rm rel}=0.0001$. They arrive at $n_{\rm H}~\lsim 0.04$~cm$^{-3}$, which our model M3 confirms if we artificially increase $V_1$ and $\epsilon_B$, and decrease $\epsilon_{\rm rel}$ to similar values to those of \cite{2019ApJ...872..191S}. In our analysis, what is new is that we tie $n_{\rm H}$ to a modeled ejecta structure with $n$ added for the outermost ejecta. We also note here that the low values of $n_{\rm H}$ found in the radio modelling of SN 1885A matches well with observations that find the central region of M31 devoid of gas (\citealt{2009ApJ...695..937B};  \citealt{2011A&A...536A..52M}).

Although not shown in Fig.~\ref{fig:1885Amod}, the 6.2 GHz luminosity at 127 years ($< 8.43\EE{21}$~erg s$^{-1}$ Hz$^{-1}$) is similar to that in model M7 (which has $n=8$) if we change from $p=2.3$ to $p=2.6$. In this $n=8$ model $n_{\rm H} = 0.072$~cm$^{-3}$ and $E_{51} = 1.13$ (i.e., close to that in the 5p02822.16 model). The choice of $p=2.6$ is motivated by the $p$-value for the spatially integrated emission in G1.9+0.3 \citep[][cf. Sect. 3.1.3.]{2020MNRAS.492.2606L}.
In summary, with $M = 1.4$ M$_{\odot}$, $p=2.3$, and our preferred values of $\epsilon_{\rm rel}$ and $\epsilon_B$, we do agree with \cite{2019ApJ...872..191S} on the upper limit of $n_{\rm H}~\lsim 0.04$~cm$^{-3}$, but for the larger value of $p=2.6$, $n_{\rm H}~\lsim 0.07$~cm$^{-3}$. For these two choices of values for $p$, $n~\gsim 9$ and $n~\gsim 8$, respectively. The latter model can accommodate $E_{51}  = 1.1$, which agrees with the kinetic energy in the 5p02822.16 model \citep{2002ApJ...568..791H}, but our limit on $E$ is only $\approx 10\%$ lower for $p=2.3$ and $n~\gsim 9$. 

\subsubsection{X-rays} \label{sec:x-rays}

\begin{table}
\centering
\caption{SN 1885A: ion density, temperature (at 123 years) and filling factor for shocked ISM and supernova ejecta} 
\label{tab:Xray_models}
\begin{tabular}{cllcccc} 
\hline
 $n$ (Model$^a$)   &  $n_{\rm ion,1}^b$  & $n_{\rm ion,2}^c$ &  $T_{1,123}$              & $T_{2,123}$             &  $\xi_{V,1}$  & $\xi_{V,2}$    \\
          &       cm$^{-3}$       & cm$^{-3}$          & $10^8$ K                  &  $10^8$ K              &                     &                       \\      
\hline
7   (M1)         & 0.682       &  0.0376                        &  13.4   &  8.19  &  0.64            &  0.87 \\
8   (M7)        & 0.318         &  0.0283                        &  14.8   &  6.22  &  0.72            &  0.91 \\
9   (M8)         & 0.155    & 0.0204                        &  16.0   &  4.91  &  0.77            &  0.96 \\
10   (M3)       & 0.0823         &  0.0151                        &  17.1   &  3.98  &  0.81            &  0.94 \\
\hline
\end{tabular}
\RaggedRight
\tablecomments{
\begin{flushleft}
    $^{\rm a}$Models are described in Table~\ref{tab:an_models}.\\
    $^{\rm b}$Ion density of shocked ISM assuming $n_{\rm He}/n_{\rm H}=0.1$.\\
    $^{\rm c}$Ion density of shocked supernova ejecta assuming $n_{\rm S}/n_{\rm Si}=1$.\\
\end{flushleft}
}
\end{table}

\begin{figure}
\includegraphics[width=0.7\textwidth]{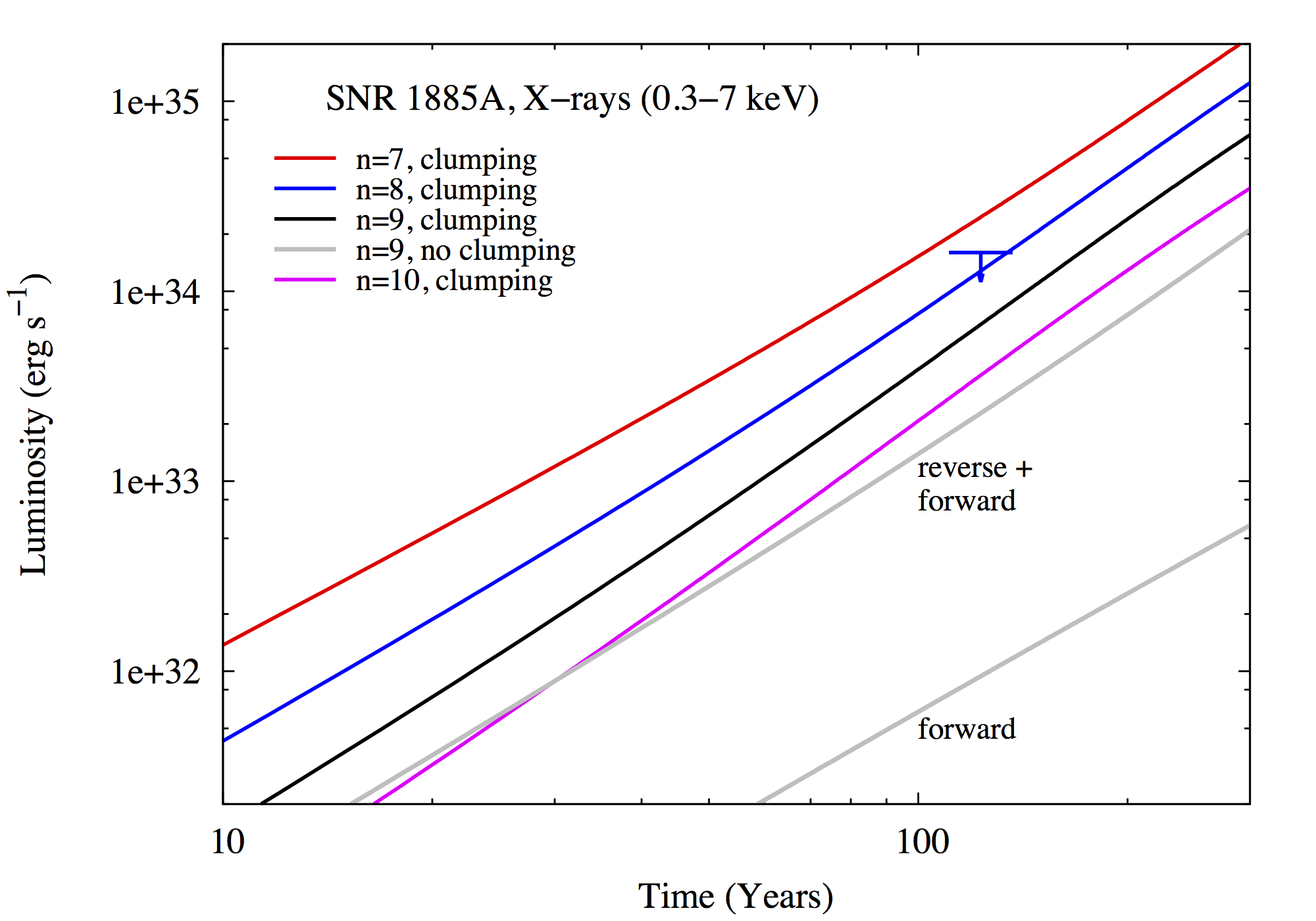}
\caption{Evolution of the thermal X-ray luminosity between $0.3-7$ keV for SN 1885A for models with values of $n$ between $7-10$ (see Table~\ref{tab:Xray_models}).
The uppermost grey line is for model M8 with emission coming from both the forward and reverse shocks. The forward shock contribution is shown by the lowermost grey line. For the models shown by the other lines clumping of the shocked supernova ejecta has been assumed. 
The upper limit is for a luminosity of $ 1.6 \times 10^{34} \ergs $ at 123 years. See text for further details.}
\label{fig:1885A_Xrays}
\end{figure}

\cite{2013A&A...555A..65H} obtained 28 X-ray counts in the range $0.2-10$ keV at the position of SN 1885A, extracted from a 1.02 Ms merged Chandra image. They note that this is a $2.6\sigma$ significance but not a clear detection for SN 1885A. We have used an average age of 123 years for SN 1885A for this data. 
Source counts were also given for the nearby remnants Braun 80, Braun 95, and Braun 101 from the same image with the counts 446, 731 and 856, respectively. 
The respective significance of the source detections are $3.2\sigma$, $4.5\sigma$, and $6.7\sigma$. In Table 1 (available at the CDS via VizieR), \cite{2013A&A...555A..65H} translate source counts into luminosity for Braun 101, and obtain $6.2 \pm {0.49} \times 10^{35}\ergs$ for a presumed spectrum of $L_{\nu} \propto \nu^{-0.7}$, a column
density of foreground material $N_H = 6.6 \times 10^{20}~{\rm cm}^{-2}$, and a distance to the remnant of 780 kpc . For similar spectra and column densities for the other 
remnants, this would mean $0.2-10$ keV luminosities of $\sim 3.2 \times 10^{35}\ergs$,  $\sim 5.3\times 10^{35}\ergs$, and $\lsim 2.0\times 10^{34}\ergs$ for Braun 80, 
Braun 95 and SN 1885A, respectively. Here we have treated the $2.6\sigma$ tentative detection for SN 1885A as an upper limit. 

The $0.2-10$ keV luminosities for Braun 95 and Braun 101 can be compared with the $0.3-7$ keV luminosities estimated by \cite{2003ApJ...590L..21K} who obtained $8^{+7.0}_{-2.2}\times10^{35}\ergs$ and $4.4^{+1.3}_{-2.2}\times10^{35}\ergs$ for Braun 95 and Braun 101, resepectively, from 37.7 ks Chandra  Advanced CCD Imaging Spectrometer data. 
They assumed 0.26~keV (Braun 95) and 0.25~keV (Braun 101) thermal spectra and $N_H = 10^{21}~{\rm cm}^{-2}$, and for Braun 90 they included a power-law component with $L_{\nu} \propto \nu^{-\alpha_{\nu}}$, with $\alpha_{\nu}=1.16^{+1.39}_{-2.16}$.  
For Braun 101 they also analyzed 50 ks of $0.3-7$ keV Chandra  High Resolution Camera data and estimated $(7.1 \pm {2.5}) \times 10^{35}\ergs$ for the same spectral model and $N_H$ as for the
37.7 ks data.
Considering the spectral range and assumed spectral shape by \cite{2013A&A...555A..65H}, their $0.2-10$ keV luminosities would be $\sim 26\%$ lower if limited to the $0.3-7$ keV range, i.e., $\sim 4.9\times 10^{35}\ergs$ for Braun 101.
This is compatible with the results of \cite{2003ApJ...590L..21K}, whereas for Braun 95 the agreement is worse. There are two XMM-Newton studies that detect Braun 101 as well (see Table \ref{tab:Xray_flux_snrs}). While a luminosity of $ 6.0\times 10^{35}\ergs$ reported for Braun 101 in \cite{2011A&A...534A..55S} for $0.2-4.5$ keV matches well with the luminosity of $ 6.2\times 10^{35}\ergs$ found by \cite{2013A&A...555A..65H}, a much higher luminosity of $ 1.2\times 10^{36}\ergs$ for the $0.35-2$ keV range is reported in \cite{2012A&A...544A.144S}, which seems to be an outlier in comparison.

For SN 1885A, the $0.3-7$ keV luminosity would be $\lsim 1.6\times 10^{34}\ergs$ (if it is 26\% lower than for the $0.2-10$ keV range), and we have marked this upper limit in Fig.~\ref{fig:1885A_Xrays}. The X-ray emission from SN 1885A can be thermal and/or non-thermal. For example, both G1.9+0.3 and Tycho have a combination of both (e.g., \citealt{2013ApJ...771L...9B}; \citealt{2023ApJ...951..103E}), but with a dominance of non-thermal synchrotron emission, although inverse Compton scattering could possibly contribute for G1.9+0.3 \citep{2024MNRAS.527.1601V}. 
In particular, G1.9+0.3, which is of similar age as SN 1885A, had in 2009 a total $1-7$ keV luminosity of $\sim 2.4\times10^{34}\ergs$ (evaluated from \cite{2011ApJ...737L..22C} and a distance of 8.5 kpc to the remnant), i.e., close to our likely upper limit for SN 1885A. A detailed comparison between SN 1885A and G1.9+0.3 is provided in Section \ref{g1.9}.

To judge whether thermal emission could play an important role for SN 1885A, we note that it is set by the temperature, density, and abundances of the shocked ejecta and shocked ISM. The temperature is given by 
\begin{equation}
	T = 2.27\times10^{9} \mu_s V_4^2~{\rm K}, 
\label{eq:temp_sh}
\end{equation}
where $\mu_s$ is the mean molecular weight per particle (for both electrons and ions), and $V_4$ the shock velocity in $10^4 \kms$. 
For fully ionized plasma with $n_{\rm He}/n_{\rm H}=0.1$,
$\mu_s = 0.61$, which applies to the shocked ISM. For the shocked ejecta (guided by the explosion model  5p02822.16 by \citealt{2002ApJ...568..791H}), we assume that they only consist of silicon and sulphur with equal number densities, which results in $\mu_s = 1.88$ for fully ionized plasma. The shock temperatures of the forward and reverse shocks at 123 years are given in Table~\ref{tab:Xray_models} for models with $a=0$ and $n$ in the range $7-10$. In the same table we also give the number densities of the ions of the shocked media ($n_{\rm ion,1}$ and $n_{\rm ion,2}$) where we have used ratio of $\rho_2/\rho_1$ (i.e., the reverse shock to forward shock density ratio) from the similarity solutions of \cite{1982ApJ...258..790C}.

To calculate the thermal X-ray emission, we assume that the density is constant within the volumes of shocked ISM and shocked ejecta, which we name $\mathbb{V}_1$ and $\mathbb{V}_2$, respectively. Similarity solutions \citep{1982ApJ...258..790C} show that the density is not really constant, and we have therefore introduced the filling factor $\xi_{V,2}$ so that the amount of swept-up mass for the ejecta, $M_2$, agrees with the similarity solutions, i.e., $M_2 = \rho_2  \xi_{V,2} \mathbb{V}_2$, and likewise for the forward shock $M_1 = \rho_1  \xi_{V,1} \mathbb{V}_1$. As can be seen from Table~\ref{tab:Xray_models} the filling factors are close to unity especially for the shocked ejecta, which is the region which is likely to dominate the thermal X-ray emission (see below). For the temperature, we presume that also this is constant within the shocked regions, but we have chosen to fix the electron temperature, $T_e$, at $20\%$ of the shock temperature, in line with, e.g., models for the late X-ray emission of SN~1993J (\citealt{2009ApJ...699..388C}; \citealt{2019ApJ...875...17K}) and for the Tycho SNR \citep{2002ApJ...581.1101H}. To numerically calculate the thermal X-ray emission we utilize the plasma code used in \cite{2019ApJ...875...17K} and described in some detail in \cite{2004AstL...30..737S}. It includes the elements H, He, C, N, O, Ne, Na, Mg, Al, Si, S, Ar, Ca, Fe, and Ni. For the shocked ISM we assume $n_{\rm He}/n_{\rm H}=0.1$ and for the other elements we choose solar abundances as given by \cite{2022A&A...661A.140M}. 

In Figure~\ref{fig:1885A_Xrays} we show the evolution between $10-300$ years of the thermal X-ray luminosity between $0.3-7$ keV 
 for models with $n \in [7,10]$ and $a=0$. The upper gray 
line is for model M8 (i.e., $n=9$) and shows the dominance of the reverse shock over the forward shock (portrayed by the lower gray line). The electron temperature of the reverse shock in this model 
is $\approx 8.5$ keV, and the $0.3-7$ keV luminosity is almost an order of magnitude below the observed X-ray upper limit. Recapitulating from Section 3.1.1, this model (with $p=2.3$) gives a radio luminosity at 
6.2 GHz that is close to the observed upper limit at that frequency (cf. Fig.~\ref{fig:1885Amod}). 

In order for thermal emission within the 
$0.3-7$ keV range to be more efficient one can invoke clumping, which could be the result of inhomogeneous ejecta. For the four models with clumping in 
Fig.~\ref{fig:1885A_Xrays} we have assumed that for one-sixth of $\xi_{V,2} \mathbb{V}_2$ the shocked ejecta are a factor of $8/3$ denser than $n_{\rm ion,2}$ in 
Table~\ref{tab:Xray_models} and that the remaining five-sixths of $\xi_{V,2} \mathbb{V}_2$ has a density that is four times lower than in the clumps. This preserves the
mass $M_2$ compared to the non-clumped models in Table~\ref{tab:Xray_models}. The shocked ejecta are assumed to be in pressure equilibrium so that the temperature 
changes inversely with density. For the $n=9$ model (black line in Fig.~\ref{fig:1885A_Xrays}) this results in a factor of $\approx 3$ higher luminosity at 123 days compared 
to the model with no clumping, but still a factor $\sim 2.4$ lower than the observed upper limit. So, clumping even by this amount is fully compatible with our estimated upper 
limit of luminosity within the $0.3-7$ keV range. If we were to treat the Chandra data as a tentative $2.6\sigma$ detection our estimates for thermal emission 
indicate that non-thermal emission is expected to dominate even if there would be significant clumping of the ejecta. However, if we allow for $p=2.6$, also the model with $n=8$ (i.e., M7)
can pass the observational tests both for radio and X-rays, even in the case of some clumping of the shocked ejecta. 

Models for thermal emission from SN 1885A were also made by \cite{2011ApJ...730...89P}. They assume $n_{\rm H}=0.1 \cm3$ and $T_2 = 0.4$ keV at 125 years. This temperature is more than an order lower than in our models, and they obtain luminosities in excess of $10^{36}$ erg~s$^{-1}$. We can reproduce this for our Model M7 (with no clumping) if we change to $T_2 = 0.4$ keV, assume solar abundances with hydrogen depressed by a factor of $10^3$ and helium by a factor of $10^2$, and set $n_{\rm ion,2}=1.0 \cm3$. \cite[Elemental abundances are, unforunately, not described by][]{2011ApJ...730...89P}. However, luminosities from thermal X-ray emission in excess of only a few per cent of $10^{36}$ are at odds with the observed upper limit of SN 1885A and the observed value for G1.9+0.3. This shows the importance of using realistic temperatures and abundances.

\subsubsection{Comparison to G1.9+0.3} \label{g1.9}
G1.9+0.3 is between $100-150$ years old \citep{2020MNRAS.492.2606L}, and thus about as old as SN 1885A. Moreover, these two remnants likely both stem from Type Ia SNe. This makes a comparison between them relevant. As noted in Section \ref{sec:x-rays}, the $1-7$ keV range luminosity for G1.9+0.3 is $\sim 2.4\times10^{34}\ergs$, and the upper limit for the $0.3-7$ keV range for SN 1885A is $\sim 1.6\times10^{34}\ergs$. In radio, the upper limits for SN 1885A are $\sim 3.5\EE{23}$~erg s$^{-1}$ Hz$^{-1}$ at 150 MHz at 133 years, and $\sim 8.4\EE{21}$~erg s$^{-1}$ Hz$^{-1}$ at 6.2 GHz at 127 years. This can be compared with the luminosities in \cite{2020MNRAS.492.2606L} for G1.9+0.3, which are $\approx 3.9\times10^{23}$~erg s$^{-1}$ Hz$^{-1}$ at 150 MHz and $\approx 1.9\times10^{22}$~erg s$^{-1}$ Hz$^{-1}$ at 6.2 GHz. Here we have interpolated between the luminosities given by \cite{2020MNRAS.492.2606L} for 5.0 GHz and 9.0 GHz, and used a distance of 8.5 kpc to G1.9+0.3. G1.9+0.3 thus seems more luminous in both radio and X-rays than SN 1885A, and it could well have been detected both in X-rays and radio (both at at 150 MHz and 6.2 GHz) had it been placed at the same distance as SN 1885A, with the clearest detection at 6.2 GHz. 
 
 In our models for SN 1885A we have assumed spherical symmetry. This agrees with the conclusion of \cite{2017ApJ...848..130F} that the distribution of Ca and Mg does not support the picture of an asymmetric explosion of SN 1885A. This may also exclude a very asymmetric distribution of the ISM in the immediate vicinity of the supernova. On the contrary, G1.9+0.3 has more of a bipolar structure in X-rays, while the strongest radio emission occurs roughly orthogonal to this bipolar structure. The inferred maximum expansion along the bipolar axis is $13\,600\pm3100\kms$ in radio \citep{2020MNRAS.492.2606L}, consistent with
 $\sim 15\,000\kms$ in X-rays \citep{2017ApJ...837L...7B}. The estimated value of $n_{\rm H}$ along this axis is $\sim 0.02\cm3$ \citep{2011ApJ...737L..22C}. In the north-west direction, where the strongest radio emission occurs, the maximum average velocity is only $\sim 8000\kms$ in radio \citep{2020MNRAS.492.2606L}, and with shock speeds even less in X-rays \citep{2017ApJ...837L...7B}. This could be explained by expansion into an inhomogeneous ISM (e.g., \citealt{2017ApJ...837L...7B}), an argument that is strengthened by a molecular cloud found close the remnant by \cite{2023PASJ...75..970E}. The bipolar structure could also be point-symmetric, and possibly signal that the supernova progenitor exploded in a planetary nebula \citep{2024RAA....24a5012S}. An asymmetric remnant structure could also be due to the orientation of the ambient magnetic field, as this affects the efficiency of how ions and electrons are accelerated round the remnant. This was was discussed for SN 1006 by \cite{2013AJ....145..104R} and \cite{2014ApJ...783...91C}.
 
 The asymmetries for G1.9+0.3 influence the comparison to SN 1885A. For example, the radio surface brightness at 1.365 GHz \citep{2017ApJ...837L...7B} is $\sim 3-4$ times higher in the north-west direction of G1.9+0.3 compared to other areas, and the area of this region is $\sim 3-4$ times smaller than the overall region of radio emission. This boosts the total emission by a factor of $\sim 2$ compared to a spherically symmetric case with a surface brightness similar to that from regions other than in the north-east. The fastest expansion of G1.9+0.3 is $30-50 \%$ faster than in our models of SN 1885A. If we assume
$p=2.6$ (which is the global average found by \citealt{2020MNRAS.492.2606L}), a homogeneous ISM with $n_{\rm H} = 0.016$, the microphysics parameters $\epsilon_B=0.04$ and $\epsilon_{\rm rel}=0.001$, 
$n=9$, $a=0$, $V_{\rm b} = 11\,200 \kms$, and $V_{\rm ej,max}({\rm 128~yrs}) =17\,500 \kms$, the modeled radio luminosity at 6.2 GHz (at 127 years) is $\approx 1.2\times10^{22}$~erg s$^{-1}$ Hz$^{-1}$, i.e., $\sim 60\%$ of the observed luminosity. For this model, the rate of luminosity increase is $\sim 0.6\%$ per year. Removing emission corresponding to the north-western region and adding the strong observed emission from that field yields a luminosity that is similar to the total overall observed emission. In this model, the forward shock advances at just below $14\,000 \kms$ in the bipolar direction, similar to the finding of \cite{2011ApJ...737L..22C}, but the model has an explosion energy of $E_{51}= 1.7$, which is higher than expected for a Type Ia SN. A solution to this could be interaction with a density enhancement in the bipolar direction that may have started only some decades ago, as suggested by \cite{2024RAA....24a5012S}. In this case the density inside this density enhancement could be $n_{\rm H} < 0.016 \cm3$ (for $n=9$) and $E$ could be closer to the canonical $10^{51}$ erg. The scenario with density enhancements is also likely to result in faster luminosity rate increase than in our homogeneous model. The observed rate is about twice as high as the $\sim 0.6\%$ per year in our model (cf. \citealt{2008MNRAS.389L..23M}). 

It appears as if all asymmetries in G1.9+0.3 make it a more efficient radio emitter (and presumably also in X-rays) than the seemingly more spherically symmetric SN 1885A. For a comparison, \cite{2019ApJ...872..191S} used  a spherically symmetric model for G1.9+0.3 and arrived at $n_{\rm H} \approx 0.18 \cm3$. The main differences compared to us is that they assumed $p=2.2$, a shock speed of $\sim 10\,000 \kms$, and $\epsilon_{\rm rel}=10^{-4}$. Their estimated density is probably more representative of that in the north-eastern region than for other parts of the remnant. The value $p=2.2$ corresponds to the average spectral index, whereas $p=2.6$ corresponds to the luminosity-weighted spectral index \citep{2020MNRAS.492.2606L}. A more detailed modeling for G1.9+0.3 should take the spatial variation of $p$ into account.

\subsection{Braun 80, Braun 95 and Braun 101}
\label{sec:remnants}

\begin{figure}
\centering
\includegraphics[width=0.6\linewidth]{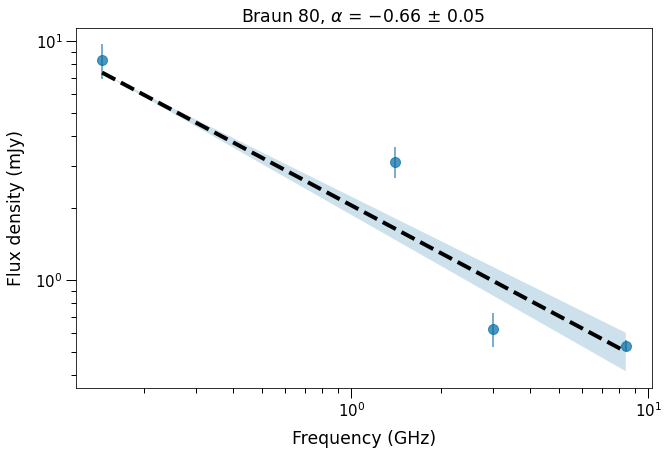} 
\caption{Spectra of Braun 80 }
\label{fig:b80_alpha}
\end{figure}

\begin{figure}
\centering
\includegraphics[width=0.6\linewidth]{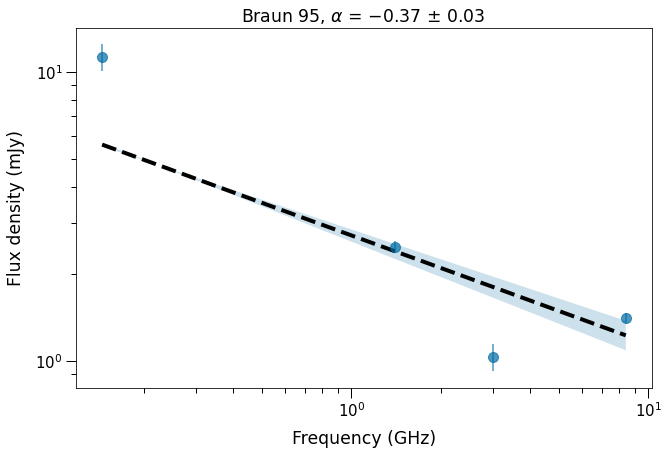}
\caption{Spectra of Braun 95}
\label{fig:b95_alpha}
\end{figure}

\begin{figure}
\centering
\includegraphics[width=0.6\linewidth]{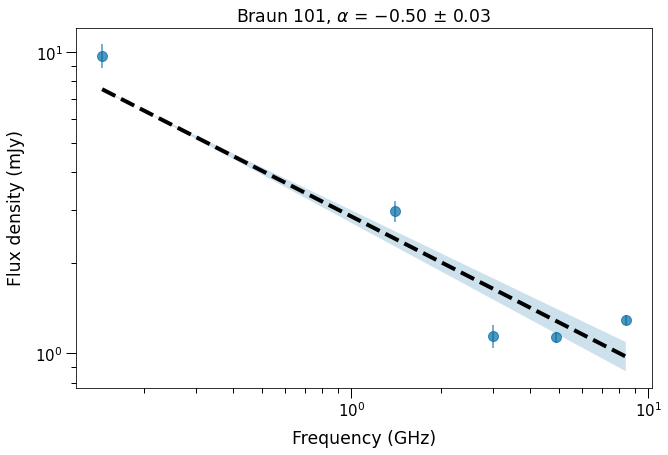}
\caption{Spectra of Braun 101.}
\label{fig:101}
\end{figure}

As seen in Figs. \ref{fig:m31lofar} and \ref{fig:m31sband} we detect the SNRs Braun 80, 95 and 101 in the LOFAR image at 150 MHz and in the VLA image at 3 GHz.
In radio, at 3 GHz, the sizes are $6\farcs17\times5\farcs55$, $6\farcs76\times5\farcs92$, and $6\farcs24\times5\farcs52$ respectively, and at 8.4 GHz
8\farcs0, $9\farcs6\times8\farcs0$, and $11\farcs6\times8\farcs4$, respectively (Section \ref{source} and Table \ref{tab:flux}). For the three remnants, the 8.4 GHz image \citep{2001AIPC..565..433S} gives the most reliable flux density among the frequencies used to estimate the spectral index. It has a good 0\farcs2 resolution and is deeper (low noise of 4 $\mu$Jy/beam), which explains the slightly larger sizes of the remnants at this frequency. This can be compared with the X-ray diameters found by \cite{2013A&A...555A..65H}, which are 7\farcs8, 8\farcs7, and 7\farcs9, respectively. The radio and X-ray luminosities are summarized in Table \ref{tab:flux} and \ref{tab:Xray_flux_snrs} (see also Section \ref{sec:x-rays}). Braun 80 has low surface brightness similar to the structure seen in other radio studies. Similar to what has been reported at 4.9 and 8.4 GHz by \cite{2003ApJ...590L..21K}, Braun 101 has a northeastern side that is 3 times brighter than the southwestern side (0.47 v 0.16 mJy) in the 3 GHz image. Braun 95 has a very apparent shell-like structure.

To estimate the properties of the remnants we use the empirical expression of \cite{2022Univ....8..653L} for the radio luminosity at 1.4 GHz
\begin{equation}
\begin{split}
	L_{1.4} \approx  2.3\times10^{24}  \left(\frac{n_{\rm ISM}}{1~\cm3}\right)^{0.85} \left(\frac{R_s}{10~{\rm pc}}\right)^{2.1} \left(\frac{V_s}{500 \kms}\right)^{1.3} \\
	{\rm erg}~{\rm s}^{-1}~{\rm Hz}^{-1}.
\end{split}
\label{eq:L_Leahy}
\end{equation}
If the remnants are in the Sedov phase, their radii and shock velocities can be described by 
\begin{equation}
	R_{s} \approx 12.5~t_4^{0.4}E_{51}^{0.2}~n_{\rm ISM}^{-0.2}~{\rm pc}
\label{eq:Sarbad_Rs}
\end{equation}
\begin{equation}
	V_{s} \approx 490~ \left(\frac{R_{s}}{12.5~{\rm pc}}\right) ~t_4^{-1}\kms,
\label{eq:Sarbad_Vs}
\end{equation}
respectively \citep{2017MNRAS.464.2326S}. Here, $t_4$ is the age in $10^4$ years. For $E_{51}=1.0$ and $n_{\rm ISM} = 0.1 \cm3$ \citep[as guided by][]{2009MNRAS.397..148L}, and radii inserted from the major axes of the 8.4 GHz images, Eqns~\ref{eq:Sarbad_Rs} and \ref{eq:Sarbad_Vs} give $t_4 \approx 0.52~(0.81, 1.31)$ and $V_{s} \approx 1150, (880, 660)\kms$ for Braun 80 (Braun 95, Braun101), where the age is $\propto (n_{\rm ISM}/E)^{1/2}$. Including SN 1885A and possibly undetected supernovae, this would indicate one supernova explosion every $\lsim 3000~(n_{\rm ISM}/0.1)^{0.5}~(E_{51,{\rm m}})^{-0.5}$ years in this central 0.5~kpc $\times$ 0.6~kpc region, where $E_{51,{\rm m}}$ is the reciprocal mean of the kinetic energies of the remnants in $10^{51}$~erg. 

We can now use Eqn~\ref{eq:L_Leahy} to estimate the expected radio luminosity at 1.4 GHz from the derived values for $R_s$ and $V_s$, as well as the assumed  value for $n_{\rm ISM} = 0.1 \cm3$, and we find $L_{1.4} \approx 2.3~(2.4,~2.5)\times10^{24}~{\rm erg}~{\rm s}^{-1}~{\rm Hz}^{-1}$ for Braun 80 (Braun 95, Braun 101). In \cite{2022Univ....8..653L} there are 31 out of 58 SNRs with shock velocities between $600-1200~\kms$ and known luminosity at 1.4 GHz, and their median value of $L_{1.4}$ is $\approx 1.0\times10^{24}~{\rm erg}~{\rm s}^{-1}~{\rm Hz}^{-1}$.
This can be compared with the observed luminosities at 1.4 GHz for Braun 80 (Braun 95, Braun 101) which are $L_{1.4} \approx 2.3~(1.8,~2.2)\times10^{24}~{\rm erg}~{\rm s}^{-1}~{\rm Hz}^{-1}$. 
In addition, from the derived spectral indices in Figures \ref{fig:b80_alpha}, \ref{fig:b95_alpha} and \ref{fig:101}, the estimated luminosities at 1.4 GHz for Braun 80 (Braun 95, Braun 101) are $L_{1.4} \approx 1.2~(1.7,~1.8)\times10^{24}~{\rm erg}~{\rm s}^{-1}~{\rm Hz}^{-1}$. 

The VLA observations are at much better resolutions compared to the LOFAR image, and all three remnants are resolved in all other frequencies except at 150 MHz. The spectral indices estimated in Figures \ref{fig:b80_alpha}, \ref{fig:b95_alpha} and \ref{fig:101} have some caveats because the scales covered by these images at different resolutions may not match very well. The 3 GHz VLA image covers scales from \SI{0.5}{\arcsecond} to \SI{18}{\arcsecond}, the 8.4 GHz images covers \SI{0.2}{\arcsecond} to \SI{20}{\arcsecond} and the 150 MHz image covers \SI{6}{\arcsecond} to several arcminutes. The 1.4 GHz image has a lower resolution (\SI{5}{\arcsecond}) and is also expected to have some confusion from the extended background.
At 8.4 GHz, the fluxes reported for these remnants are only partially recovered in \cite{2001AIPC..565..433S} because large scale diffuse emission greater than \SI{10}{\arcsecond} are filtered out, as explained in \cite{2005xrrc.procE4.16S}. 
Given these caveats, the spectral indices of the remnants (\textminus 0.66$\pm$0.05, \textminus 0.37$\pm$0.03 and \textminus 0.5$\pm$0.03 for Braun 80, 95 and 101 respectively) match fairly well with those found in previous studies (\citealt{2003ApJ...590L..21K} and \citealt{2001AIPC..565..433S}). \cite{2003ApJ...590L..21K} mention that  the flux density measured at 1.4 GHz might be confused with the extended background but derive a spectral index of \textminus 0.77 for Braun 80. \cite{2001AIPC..565..433S} find spectral index of \textminus 0.3 and \textminus 0.5 for Braun 95 and 101 respectively while adopting a nominal \textminus 0.5 for Braun 80 due to the confused flux densities. As noted in the captions, 3\% errors were used for the 8.4 and 4.9 GHz fluxes to estimate the spectral indices. This is because only the noise level is reported in \cite{2001AIPC..565..433S} and \cite{2003ApJ...590L..21K}. We note here also that the indices are sensitive to these error assumptions. For example, for 1\% error assumed in addition to the 3$\sigma$ noise, the indices then become \textminus 0.64$\pm$0.05, \textminus 0.3$\pm$0.03 and \textminus 0.1$\pm$0.03 for Braun 80, 95 and 101, respectively. The difference for Braun 101 is due to the dominance of the 4.9 GHz and 8.4 GHz measurements in this case.

\section{Conclusion} \label{conc}
We presented the first LOFAR image of the centre of M31 at a resolution of \SI{6}{\arcsecond} and rms of 0.1 mJy/beam at a 150 MHz. Along with this, we used a 3 GHz VLA image with a beam size of \SI{0.5}{\arcsecond} and rms of 3 \muJ/beam, and previously published radio and X-ray data to study four remnants in this central region of M31 (SN 1885A, Braun 80, 95 and 101).

In our models for the historical SN 1885A, we assume spherically symmetric ejecta that expand homologously and density structure that can be approximated by two power-laws; $\rho_i (V,t)\propto V^{-a}t^{-3}$ and $\rho_o (V,t) \propto V^{-n}t^{-3}$, where $a$ and $n$ correspond to the slopes of the inner and outer ejecta structure, respectively. Using  values guided by previous studies of SN 1885A, and the similarity solutions of \cite{1982ApJ...258..790C}, we model the SN ejecta for $a$-values 0 and 1 and $n$-values 7 -- 12 to approximate the full density profile. We estimate time $t_{\rm b}$ which gives the age at which SN 1885A is expected to transition to the Sedov-Taylor stage. We find larger values of $t_{\rm b}$ than previously estimated, indicating that the remnant will stay in the free-expansion phase for at least another couple of centuries.

We also perform radio modelling using our upper limit at 150 MHz and the upper limit at 6.2 GHz in \cite{2019ApJ...872..191S}. We use $\epsilon_{\rm rel}=0.001$ and $\epsilon_B=0.01$ which are close to the geometric means of SNRs Tycho, Kepler, and G1.9+0.3. We also assume the spectral index for the power law distribution of synchrotron-emitting electrons to be $p=2.3$, which is within the range typical for SNRs. For models with $n \in [7,9]$, we obtain post-shock magnetic field energy density values of $B$ similar to $B = 229~\mu G$ which is the geometric mean for the remnants Tycho, Kepler and G1.9+0.3. For models with $n \in [10,12]$ $B$ is closer to $100~\mu G$ which is similar to the $B$-values found for RX~J1713.7-3946, RCW~86 and SN~1006.  The 6.2 GHz data \citep{2019ApJ...872..191S} is the deepest radio image obtained at the position of SN 1885A and is the most constraining for our models. Emission at 6.2 GHz would have been observed if similar to our models M1, M2 and M8, and detectable at 150 MHz if similar to the models M1 and M2. For a spectral index of $p=2.3$ for the relativistic electrons, we agree with \cite{2019ApJ...872..191S} on the upper limit of $n_{\rm H}~\lsim 0.04$~cm$^{-3}$. For the steeper index $p=2.6$, the limit is $n_{\rm H}~\lsim 0.07$~cm$^{-3}$. For these two choices of values for $p$, $n~\gsim 9$ and $n~\gsim 8$, respectively. The latter model can accommodate $E_{51}  = 1.1$, which agrees with the kinetic energy in the 5p02822.16 model \citep{2002ApJ...568..791H}, but our limit on $E$ is only $\approx 10\%$ lower for $p=2.3$ and $n~\gsim 9$.

\cite{2013A&A...555A..65H} report a $2.6\sigma$ significance for SN 1885A in X-ray data from deep Chandra imaging at $0.2-10$ keV. We estimate a luminosity value of $\lsim 2.0\times 10^{34}\ergs$ from their reported source count. We treat this as an upper-limit for modelling the X-ray emission and check if thermal emission would play a significant role. Even if clumping of the ejecta is invoked for more efficient thermal emission, we find that non-thermal emission is expected to dominate the X-ray emission.

We then compare our results on SN 1885A with those on G1.9+0.3. We use the luminosities at 5 GHz and 9 GHz reported by \cite{2020MNRAS.492.2606L}, and interpolate these values to 150 MHz and 6.2 GHz using a distance of 8.5 kpc for G1.9+0.3. We find that, had G1.9+0.3 been at the same distance as SN 1885A, it could very well have been detected not only at 6.2 GHz, but also at 150 MHz. While we assume spherical symmetry for SN 1885A, which is supported observationally, G1.9+0.3 has asymmetries. The observed luminosity increase is about twice as high as the 0.6\% per year seen in our spherically symmetric model. It appears likely that the asymmetries make G1.9+0.3 a much more efficient radio and X-ray transmitter. Further studies of the remnant should take into account these asymmetries using a spatially-varying $p$ value (since $p=2.2$ corresponds to an average spectral index while $p=2.6$ corresponds to a luminosity-weighted spectral index).

We use the 3 GHz image to characterise the SNRs Braun 80, 95 and 101, with Braun 80 having low surface brightness that affects its spectral index estimation, Braun 95 having 3 times more emission from the northeastern side compared to the southwestern side, and Braun 101 having a very apparent shell-like structure. We estimate two properties of the remnants, age and shock velocity using the radii of the remnants from \cite{2001AIPC..565..433S}, and assuming an explosion energy of $E_{51}=1.0$ and an ISM density of ($n_{\rm ISM} = 0.1 \cm3$). We find ages of 5200, 8100 and 13\,100 years and shock velocities of 1150, 880, 660 $\kms$ for Braun 80, 95 and 101, respectively. For this 0.5~kpc $\times$ 0.6~kpc central region of M31, this indicates that there is one supernova explosion every $\lsim 3000$ years. This result scales with the ISM density and presumed explosion energy as $\propto (n_{\rm ISM}/E)^{1/2}$. We estimate spectral indices of \textminus 0.66, \textminus 0.37 and \textminus 0.5 for Braun 80, 95 and 101, respectively, which match well with the indices found by \cite{2003ApJ...590L..21K} and \cite{2001AIPC..565..433S}.

An SNR is expected to be brighter at lower frequencies due to optically-thin power-law synchrotron emission. However, even at low frequencies of 150 MHz, SN 1885A remains undetected in the radio, as we show with LOFAR. The VLA and X-ray observations of the studied region will be difficult to push deeper with present-day facilities and techniques. However, the LOFAR data in this study can be made more sensitive using VLBI with sub-arcsecond resolution instead of the \SI{6}{\arcsecond} data used here, as we have demonstrated for SN 2011dh in \cite{2023ApJ...953..157V}. With a spectral index of \textminus 0.6, which is typical for young SNRs, the expected flux density of SN 1885A at 150 MHz could be 0.1 mJy ($\sim10$ times higher than the upper limit at 6.2 GHz). As discussed in this work, the forward shock radius is expected to be $\approx 2$ pc, meaning the size of SN 1885A in radio would be $\sim 1\farcs1$ (diameter of 4 pc). Typical VLBI observations with LOFAR with $0\farcs3$ resolution (which is about the same as for the 6.2 GHz observations, cf. \cite{2019ApJ...872..191S}) and $\sim 80~\muJ$ noise level may very well be at the limit of detection for SN 1885A. Such attempts for M31 are underway (Bonnassieux et al. in prep), and this could make the LOFAR data at least as useful as the VLA data, especially for SN 1885A.

\section*{Acknowledgements}
We are grateful to the anonymous referee for insightful comments. It is a pleasure to thank J. Moldón and M. C. Toribio for useful discussions.
This paper is based on data obtained with the International LOFAR Telescope (ILT) under project code LC10\_014. LoTSS Data Release 2 data are described by \cite{2022A&A...659A...1S}.
LOFAR data products were provided by the LOFAR Surveys Key Science project (LSKSP; \url{https://lofar-surveys.org/}) and were derived from observations with the International LOFAR Telescope (ILT). LOFAR \citep{2013A&A...556A...2V} is the Low Frequency Array designed and constructed by ASTRON. It has observing, data processing, and data storage facilities in several countries, which are owned by various parties (each with their own funding sources), and which are collectively operated by the ILT foundation under a joint scientific policy. The efforts of the LSKSP have benefited from funding from the European Research Council, NOVA, NWO, CNRS-INSU, the SURF Co-operative, the UK Science and Technology Funding Council and the Jülich Supercomputing Centre. 
MPT acknowledges financial support from the Severo Ochoa grant CEX2021-001131-S and from the National grant
PID2020-117404GB-C21, funded by MCIU/AEI/ 10.13039/501100011033.
The National Radio Astronomy Observatory is a facility of the National Science Foundation operated under cooperative agreement by Associated Universities, Inc. This research has made use of the NASA/IPAC Extragalactic Database (NED), which is funded by the National Aeronautics and Space Administration and operated by the California Institute of Technology.

\vspace{5mm}
\facilities{LOFAR, VLA}

\software{
          \textsc{APLpy} (\citealt{aplpy2012,aplpy2019})
          \textsc{NumPy} \citep{harris2020array},
          \textsc{SciPy} \citep{2020SciPy-NMeth},
          \textsc{Matplotlib} \citep{Hunter:2007},
          \textsc{CASA} \citep{2022PASP..134k4501C},
          \textsc{LINC} \citep{2019A&A...622A...5D},
          \textsc{DDFacet} \citep{2023ascl.soft05008T}
          }

\bibliography{m31}{}
\bibliographystyle{aasjournal}
\listofchanges
\end{document}